\documentclass[%
 reprint,
nofootinbib,
  showpacs,
 amsmath,amssymb,
 aps,
prb,
floatfix,
longbibliography,
superscriptaddress,
]{revtex4-2}

\usepackage[dvipdfmx]{graphicx} 
\usepackage{dcolumn}
\usepackage{pifont} 
\usepackage{bm}
\usepackage{multirow}
\usepackage{longtable}
\usepackage[utf8]{inputenc}   
\usepackage[T1]{fontenc}      
\usepackage{lmodern}          
\usepackage{csquotes}         
\usepackage{amsmath}          
\usepackage{mathtools}        
\usepackage{amstext}         
\usepackage{amssymb}          
\usepackage{subdepth}         
\usepackage{xcolor}         
\usepackage[
  separate-uncertainty=true,
  per-mode=reciprocal,
  range-phrase=--]{siunitx}  
\usepackage{braket}           %
\usepackage[pass]{geometry}  
\usepackage[normalem]{ulem}
\usepackage{times}
\usepackage[colorlinks,breaklinks,bookmarks=true,citecolor=blue,linkcolor=blue,urlcolor=blue]{hyperref}
\usepackage[capitalize]{cleveref}

\begin{document}

\preprint{APS/123-QED}

\title{
    Electronic structures of spin-orbit-coupled metal candidate PbRe$_2$O$_6$:
    one dimensionality and molecular orbital formation
}

\author{Yuki Yanagi}
\affiliation{Liberal Arts and Sciences, Toyama Prefectural University, Imizu, Toyama 939-0398, Japan}
\author{Michi-To Suzuki}
\affiliation{Department of Materials Science, Graduate School of Engineering, Osaka Metropolitan University, Sakai, Osaka 599-8531, Japan}
\affiliation{Center for Spintronics Research Network, Graduate School of Engineering Science, The University of Osaka, Toyonaka, Osaka 560-8531, Japan}

\begin{abstract}

    We present a first-principles investigation of the electronic structure of the inversion-symmetry-broken spin-orbit-coupled metal candidate  PbRe$_2$O$_6$. Our calculations reveal that the Fermi surfaces derived from the $d_{yz}$ and $d_{zx}$ orbitals exhibit pronounced one-dimensional characteristics, which naturally account for the highly anisotropic charge transport observed experimentally. In addition, the $d_{x^2-y^2}$ orbitals on each Re haxagon form molecular orbitals, where the resulting $E_g$ molecular states generate nearly dispersionless bands in close proximity to the Fermi level. The coexistence of these quasi-1D Fermi surfaces and molecular-orbital-induced flat bands provides a possible microscopic origin for the successive phase transitions observed in PbRe$_2$O$_6$.

\end{abstract}

\maketitle

\section{Introduction}

\begin{figure*}[t]
    \centering
     \includegraphics[width=\textwidth]{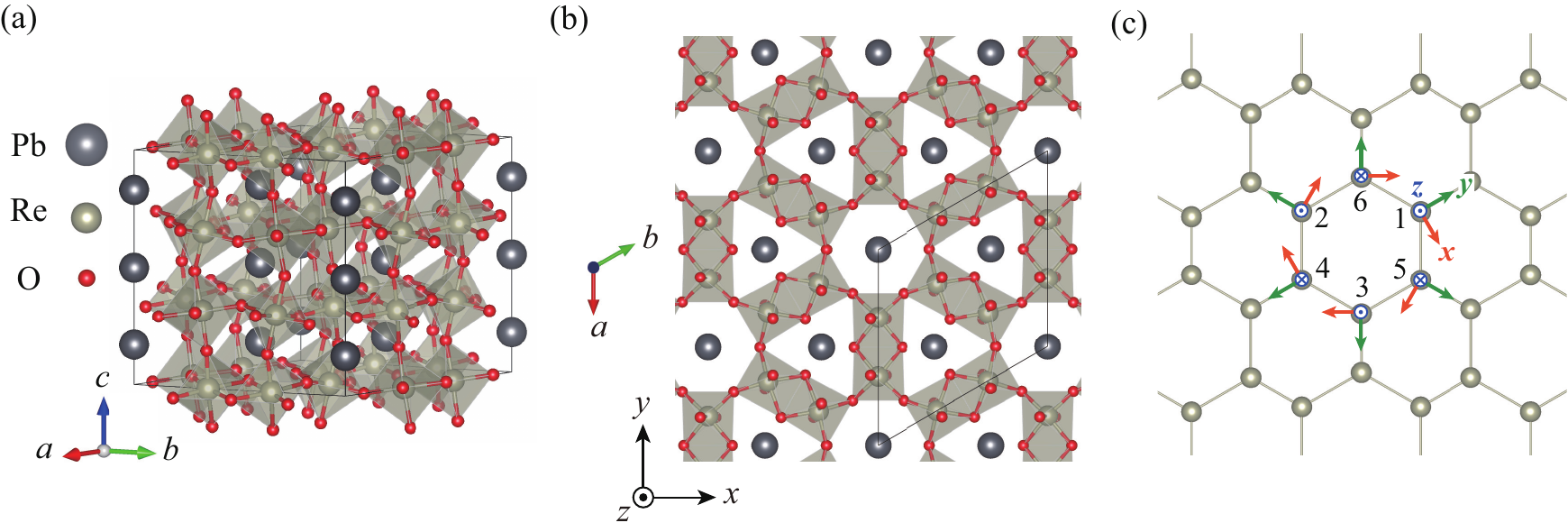}
    \caption{(a) Crystal structure of PbRe$_2$O$_6$ and (b) that on $ab$-plane. The solid line represents the conventional unit cell. (c) Same as (b) but the only Re-atoms with local coordinates are shown.}
    \label{fig:crystal}
\end{figure*}

Entangled states emerging from spin, charge, and orbital degrees of freedom give rise to a variety of fascinating physical phenomena, including cross-correlation and novel charge/spin transports~\cite{Khomskii_ChemRev2020,Nagaosa_JPSJ.92.081002,Hayami_JPSJ.93.072001}.
Metal with strong spin-orbit coupling, often referred to as spin-orbit-coupled metal, provides a fertile ground for such intriguing physics~\cite{Fu_PhysRevLett.115.026401}. Among these phenomena, spontaneous inversion symmetry breaking as a Fermi liquid instability has attracted much interest~\cite{Fu_PhysRevLett.115.026401}.
Under parity-breaking orders, the spin-degeneracy of the Fermi surfaces is lifted at general $\bm{k}$-points, resulting in characteristic spin textures in reciprocal space.

Recently, $5d$-electron compounds have been extensively studied as model systems showing interplay of strong correlation effects and spin-orbit coupling~\cite{William_annurev_2014,Takayama_JPSJ.90.062001}, such as novel Mott insulating state~\cite{Kim_PhysRevLett.101.076402,Kim_science.1167106.2009,Watanabe_PhysRevLett.105.216410,Tokura_RPP_2025}, noncoplanar magnetism~\cite{Shinaoka_PhysRevLett.108.247204,Yamaura_PhysRevLett.108.247205,Shinaoka_JPCM2019}, Kitaev physics~\cite{Jackeli_PhysRevLett.102.017205,Mazin_PhysRevLett.109.197201,Winter_PhysRevB.93.214431,Yamaji_PhysRevLett.113.107201,Singh_PhysRevB.82.064412,Kitagawa_Nature_2018,Takagi_NatRevPhys_2019,Motome_JPCM2020}, and multipolar ordering~\cite{Chen_PhysRevB.82.174440,Paramekanti_PhysRevB.101.054439,Kubo_PhysRevB.107.235134,Erickson_PhysRevLett.99.016404,Lu_NatCommun_2017,Hirai_JPSJ.88.064712,Ishikawa_PhysRevB.100.045142}.
The $5d$-electron pyrochlore compound Cd$_2$Re$_2$O$_7$ is a promising candidate for a spin-orbit-coupled metal exhibiting spontaneous inversion symmetry breaking~\cite{Yamaura_JPSJ.71.2598,Hiroi_JPSJ.87.024702}.
It crystallizes in a centrosymmetric cubic structure with space group $Fd\bar{3}m$ at room temperatures, and undergoes two successive structural phase transitions at $T_{\mathrm{s}1}\sim \qty{205}{\kelvin}$ and $T_{\mathrm{s}2}\sim \qty{112}{\kelvin}$. These transitions give rise to noncentrosymmetric structures with space groups $I\bar{4}m2$ for $T_{\mathrm{s}2}<T<T_{\mathrm{s}1}$ and $I4_122$ for $T<T_{\mathrm{s}2}$~\cite{Yamaura_JPSJ.71.2598}.
The resistivity remains metallic over the entire temperature range, while the magnetic susceptibility, which exhibits only weak temperature dependence, becomes strongly suppressed below $T_{\mathrm{s}1}$~\cite{Hanawa_PhysRevLett.87.187001,Jin_JPCM_2002,Hiroi_JPSJ.71.1634}. Theoretical studies provide possible scenarios of these parity-breaking orders~\cite{Sergienko_JPSJ.72.1607,Norman_PhysRevB.92.075113,Matteo_PhysRevB.96.115156,Hayami_PhysRevLett.122.147602}.

Another Re-based material, PbRe$_2$O$_6$, shares several key similarities with Cd$_2$Re$_2$O$_7$.
This compound crystallizes in a centrosymmetric trigonal structure with space group $R\bar{3}m$ at room temperatures~\cite{Wentzell_zaac.19855280905}, and undergoes two successive structural phase transitions at $ T_{\mathrm{s}1}\sim \qty{265}{\kelvin}$ and $ T_{\mathrm{s}2}\sim \qty{123}{\kelvin}$~\cite{TAJIMA2020121359}.
Its resistivity exhibits metallic temperature dependence, and the magnetic susceptibility shows only weak temperature variation, being noticeably suppressed across the transition at $T_{\mathrm{s}1}$~\cite{TAJIMA2020121359}.
These features, reminiscent of Cd$_2$Re$_2$O$_7$, may indicate the emergence of spontaneous inversion symmetry breaking at $ T_{\mathrm{s}1}$ and $ T_{\mathrm{s}2}$. However, the detailed crystal structures below $ T_{\mathrm{s}1}$ including the atomic positions and space groups, have not yet been fully resolved.

PbRe$_2$O$_6$ also displays several features that are markedly distinct from Cd$_2$Re$_2$O$_7$.
Most notably, it exhibits strongly anisotropic charge transport above $T_{\mathrm{s1}}$.
In this temperature region, the resistivity along the $c$-axis, $\rho_{zz}$, is much smaller than the in-plane components $\rho_{xx}=\rho_{yy}$~\cite{Hirai_p.c.}, in contrast to Cd$_2$Re$_2$O$_7$, whose electronic transport is nearly isotropic due to its cubic symmetry at room temperature.
Such strong transport anisotropy, $\rho_{zz}\ll \rho_{xx}=\rho_{yy}$, strongly suggests a one-dimensional nature of electronic structure.

In the present paper, we investigate the electronic states of PbRe$_2$O$_6$ using first-principles calculations. We demonstrate that its Fermi surfaces (FSs) exhibit pronounced one-dimensionality and identify nearly dispersionless bands near the Fermi level originating from molecular orbitals formed by the Re $d$ orbitals on a hexagonal network.
By analyzing the orbital composition of Re 5$d$ states contributing to the Bloch bands, we clarify the microscopic origins of both the quasi-one-dimensional Fermi surfaces and the flat-band formation, which could play a key role in the structural phase transitions in PbRe$_2$O$_6$.

\section{Method}

We perform the density functional calculation  using the \textsc{Wien2k} code~\cite{wien2k_2001,wien2k_2020}, where the structural parameters are extracted from the experimental values~\cite{Wentzell_zaac.19855280905}.
Crystal structures of PbRe$_2$O$_6$ are shown in Fig.~\ref{fig:crystal}.
Along the $c$-axis, ReO$_6$ octahedra are aligned with corner-shared structure and in the $a$-$b$ plane,  the (next) nearest neighbor Re-Re bond is composed of the edge-shared (corner-shared) structure as shown in Figs.~\ref{fig:crystal}(a) and \ref{fig:crystal}(b).
The muffin-tin radius for Pb, Re, and O-atoms are set to $\qty{2.35}{\mathrm{bohr}}$, $\qty{1.87}{\mathrm{bohr}}$, and $\qty{1.60}{\mathrm{bohr}}$, respectively.
The plane wave cutoff $K_{\mathrm{max}}$ is determined so as to satisfy the relation $R_{\mathrm{MT}}K_{\mathrm{max}} = 8.0$, where $R_{\mathrm{MT}}$ is the smallest muffin-tin radius.

To analyze the low energy electronic structures in detail,
we construct an effective tight-binding model using the \textsc{Wannier90} package~\cite{Pizzi_2020} through the \textsc{wien2wannier} interface~\cite{Kunes_20101888}. Octahedrally-coordinated O-atoms surrounding a Re-atom produces a large crystalline electric field splitting between Re-$t_{2g}$-orbitals ($d_{x^2-y^2(=v)}$, $d_{yz}$, $d_{zx}$) and Re-$e_g$-orbitals ($d_{3z^2-r^2}$, $d_{xy}$), where the local coordinates for six Re atoms in a unit cell are set as shown in Fig.~\ref{fig:crystal}(c).
Our minimal effective model thus includes low-lying  $t_{2g}$-orbitals on six Re-atoms per spin, i.e., $3\times 6\times 2= 36$-orbitals.

\begin{figure}[thb]
    \centering
    \includegraphics[width=1.0\linewidth]{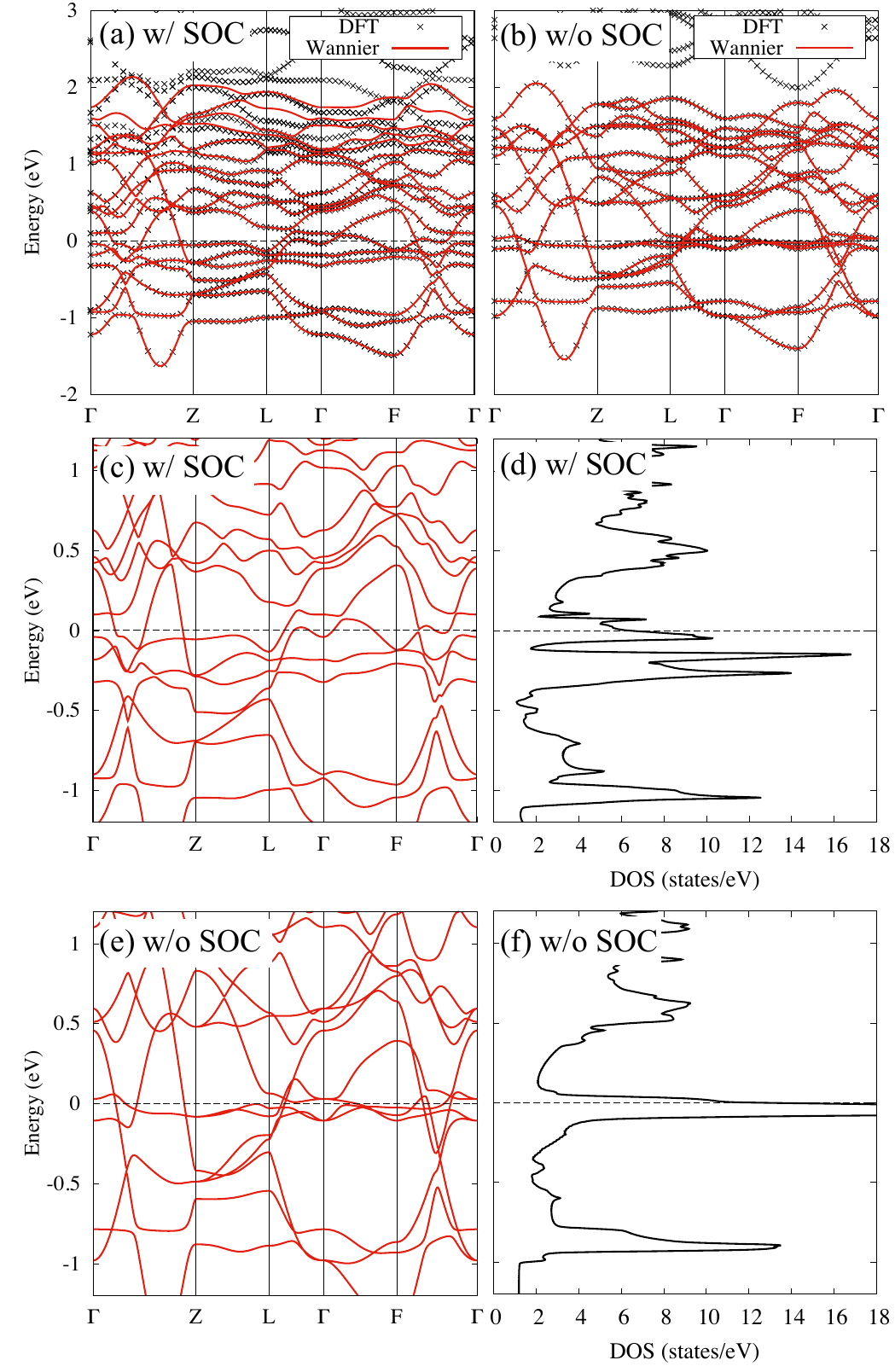}
    \caption{
        Band structures of PbRe$_2$O$_6$ (a) with and (b) without spin-orbit coupling. The solid line and cross represent the results obtained by the Wannier interpolation and density functional calculation, respectively.
        (c) Enlarged view of the low energy band structure and (d) density of states  per spin with spin-orbit coupling. (e) and (f) those without spin-orbit coupling.
    }
    \label{fig:band}
\end{figure}

\section{Results}

Figures~\ref{fig:band}(a) and \ref{fig:band}(b) show the band structures with and without spin-orbit coupling (SOC), respectively.
For both cases, The low energy states from $\qty{-2}{\electronvolt}$ to $\qty{3}{\electronvolt}$ are mainly composed of $d_v$, $d_{yz}$, and $d_{zx}$-orbitals. The band structure in 36-orbital tight-binding model thus well reproduces the first-principles one.
The enlarged view of the band structure near the Fermi energy with SOC is shown in Fig.~\ref{fig:band}(c) and the density of states (DOS) is shown in Fig.~\ref{fig:band}(d).
One can observe characteristic multi-peak structures in the DOS in the range from $\qty{-0.4}{\electronvolt}$ to the Fermi energy, as well as a sharp peak around $\qty{-1.0}{\electronvolt}$.
Correspondingly, the band structures around $\qty{-0.15}{\electronvolt}$ and $\qty{-1.0}{\electronvolt}$ are weakly dispersive as shown in Fig.~\ref{fig:band}(c).
The very large DOS just below the Fermi energy might be relevant to the electronic and/or electron-lattice instability in PbRe$_2$O$_6$.

Such characteristic features in the electronic structures are more pronounced in the SOC-free case as shown in Figs.~\ref{fig:band}(e) and \ref{fig:band}(f).
The energy bands without the SOC just below the Fermi energy and around $\qty{-1.0}{\electronvolt}$ are more dispersionless than those with the SOC and the multi-peak structures in the DOS around the Fermi energy merge into a sharp peak in the SOC-free case.
The characteristic peak structures in the DOS are well captured by the molecular orbital description derived from the $d_v$-orbitals as will be shown later.

\begin{figure}
    \centering
    \includegraphics[width=1.0\linewidth]{Fig3.pdf}
    \caption{Fermi surfaces of PbRe$_2$O$_6$  in the presence of the SOC (a) for band-11/12 and (b) for band-13/14. (c) and (d) Intersections on $k_x=0$ and $k_y=0$-planes, respectively.
        RGB color-coding represents the Re-$5d$ orbital weights as follows, R: $d_{zx}$, G: $d_{yz}$, and B: $d_v$-orbital weights.
    }
    \label{fig:FS}
\end{figure}

To obtain a deeper understanding of low energy electronic states, we demonstrate the Fermi surfaces (FSs) with orbital weights in Fig.~\ref{fig:FS}.
It is  observed that  the FSs is highly one-dimensional along $k_z$-direction especially for the 13/14th-band.
The large part of the FSs for the 11/12th-band is also nearly flat in $k_x$-$k_y$ plane, indicating the considerable one-dimensional character.
Focusing on the orbital character, the FSs of the 13/14th-bands around  the $(0,0,k_z)$ line  are mainly composed of the $d_{zx}$-orbital, while the $d_{yz}$ and $d_v$-orbital contributions are large for the FSs of the 13/14th-bands away from the $(0,0,k_z)$ line.

\begin{figure}
    \centering
    \includegraphics[width=1.0\linewidth]{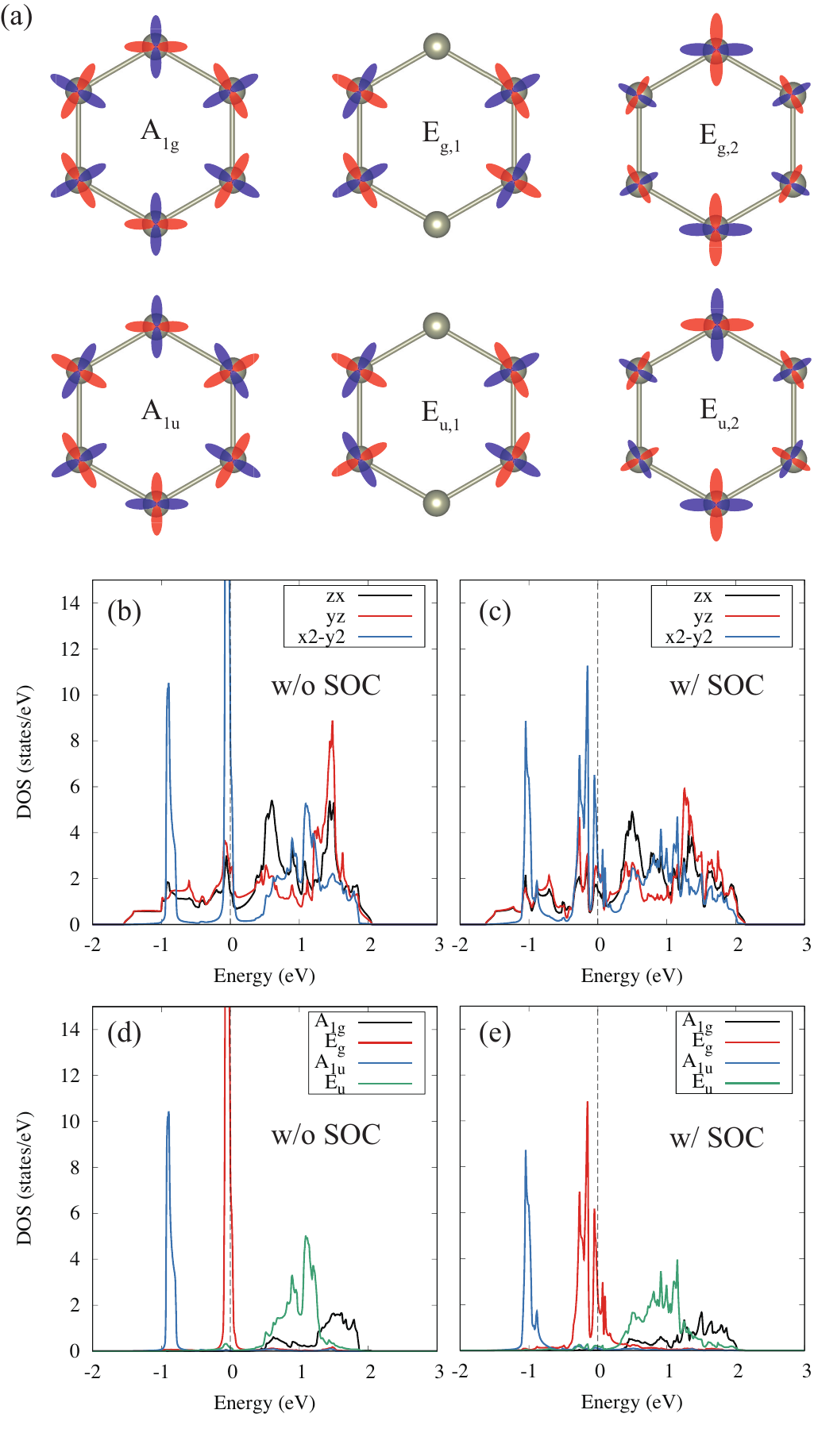}
    \caption{
        (a) Molecular orbitals derived from $d_v$-orbitals on a hexagon in the SOC-free case.
        Size of orbitals corresponds to the weight of the wave function. (b) DOS for molecular orbital representation from $d_v$-orbitals in the absence of the SOC and (c) presence of the SOC.
    }
    \label{fig:mo}
\end{figure}

Here, we demonstrate that the molecular orbital picture schematically shown in Fig.~\ref{fig:mo}(a) well describes the nearly dispersionless energy bands and the corresponding large DOS.
The orbital-resolved DOS in the absence (presence) of the SOC is shown in Fig.~\ref{fig:mo}(b) [\ref{fig:mo}(c)].
The $d_{yz}$ and $d_{zx}$-orbital components show several relatively broad peak structures.
On the other hand, $d_v$-orbital component mainly comprises the sharp peak structures around the Fermi energy and $\qty{-1.0}{\electronvolt}$ as mentioned before.
The $d_v$-orbital DOS components are well characterized by the symmetry-adapted molecular orbital.
Since the unit cell in PbRe$_2$O$_6$ contains six Re atoms,
six molecular orbitals are derived from each $d_{yz}$, $d_{zx}$, and $d_v$-orbitals in the spinless case.
For $d_{zx}$ and $d_{v}$-orbitals, molecular orbitals are decomposed into $A_{1g}$, $E_{g}$, $A_{1u}$, and $E_u$ irreducible representations (irreps.) under the crystallographic $D_{3d}$ point group, while for $d_{yz}$-orbitals, those are decomposed into
$A_{2g}$, $E_{g}$, $A_{2u}$, and $E_u$ irreps.
We schematically show symmetry-adapted molecular orbitals derived from $d_v$-orbitals in Fig.~\ref{fig:mo}(a).
Figure~\ref{fig:mo}(d) shows the DOS projected on these molecular orbitals in the SOC-free case.
The electronic states localized just below the Fermi energy and around $\qty{-1.0}{\electronvolt}$ are clearly composed of the $A_{1u}$ and $E_g$-molecular orbitals, respectively. On the other hand, the $E_u$ and $A_{1g}$ molecular orbital states are widely spread above the Fermi energy.
The molecular orbital picture is thus quite efficient for the description of the electronic states in the SOC-free case.

Figure~\ref{fig:mo}(e) shows the molecular orbital-resolved DOS in the finite SOC case. Although the molecular orbitals shown in Fig.~\ref{fig:mo}(a) are no longer irreps. due to the finite SOC,  each components are well separated in energy as in the SOC free case.
The molecular orbital picture also works well even in the finite SOC case.
The main difference in DOS of the finite SOC case from the SOC free one is the splitting of the $E_g$-components just below the Fermi level, which is a natural consequence of the effects of the SOC on $E_g$-states as follows: $E_g \otimes E_{\frac{1}{2}g} = E_{\frac{1}{2}g} \oplus E_{\frac{3}{2}g}$.

\begin{figure}[t]
    \centering
    \includegraphics[width=1.0\linewidth]{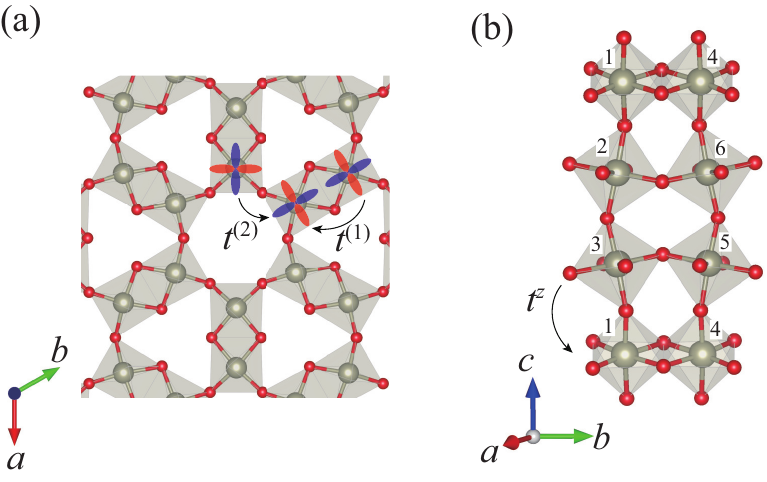}
    \caption{Schematics of (a) in-plane and (b) out-of-plane hoppings.}
    \label{fig:hopping}
\end{figure}

Let us discuss the origin of characteristic electronic structures, i.e., (1)~molecular orbital formation of $d_v$-orbitals and (2)~one-dimensionality, from the perspective of the tight-binding parameters.
As shown in Fig.~\ref{fig:hopping}(a), the $d_v$-orbitals are directed along the nearest Re-Re bonds and the nearest-neighbor hopping between  $d_v$-orbitals  $t^{(1)}_{v,v}$  thus is dominated by the $\sigma$-hopping, which is large in the ordinary cases.
If $t^{(1)}_{v,v}$  is quite larger than the hopping in a Re-hexagon  $t^{(2)}_{v,v}$ , the bonding and anti-bonding orbitals are formed on the center of the nearest Re-Re bonds.
However, the $\sigma$-hopping is quite small $t^{(1)}_{v,v}\sim\qty{-44}{\milli\electronvolt}$ in the present Wannier model as shown in Table~\ref{tab:hopping}.
To clarify the origin of the small $t^{(1)}_{v,v}$, we construct the $t_{2g}$-$p$ Wannier model, which explicitly includes the O-$2p$ orbitals in addition to the Re-$t_{2g}$ orbitals.
In the  $t_{2g}$-$p$ Wannier model, the direct $\sigma$-hopping is quite larger than the $t_{2g}$ Wannier model $t^{(1)}_{v,v}\sim \qty{-440}{\milli\electronvolt}$.
Therefore, the small $\sigma$-hopping between $d_v$-orbitals in the $t_{2g}$ Wannier model is considered to be due to the cancellation of the direct $d$-$d$ $\sigma$-hopping and the effective ones via the O-$2p$ orbitals.
Similar features are previously reported in several Kitaev candidate materials, such as Na$_2$IrO$_3$~\cite{Mazin_PhysRevLett.109.197201,Winter_PhysRevB.93.214431}, in which the nearest-neighbor Ir-Ir bond is composed of the edge-shared IrO$_6$ octahedra similar to the present case [see Figs.~\ref{fig:crystal}(b) and \ref{fig:hopping}(a)].
In contrast to the small $d$-$d$ $\sigma$-hopping $t^{(1)}_{v,v}$, the hopping between $d_v$-orbitals in a Re-hexagon $t^{(2)}_{v,v}$ is quite large $t^{(2)}_{v,v}\sim \qty{522}{\milli\electronvolt}$.
The sufficiently small $|t^{(1)}_{v,v}/t^{(2)}_{v,v}|\sim 0.08$ is a key ingredient of the molecular orbital formation derived from $d_v$-orbitals.

The one-dimensional electronic structure along $c$-axis is also naturally understood from the hopping parameters.
We show the nearest out-of-plane hopping between orbitals $\ell$ and $\ell'$, $t^z_{\ell,\ell'}$ in the rightmost column in Table~\ref{tab:hopping} [see also Fig.~\ref{fig:hopping}(b)].
When orbitals $\ell$ and $\ell'$ are directed along $c$-axis, that is, $\ell$ and $\ell'$ are $zx/yz$-orbitals, $t^z_{\ell,\ell'}$ have large value,  $|t^z_{yz,yz}|\sim \qty{328}{\milli\electronvolt}$ and $|t^z_{zx,yz}|\sim \qty{399}{\milli\electronvolt}$.
The out-of-plane hopping for $d_{zx}$- and $d_{yz}$-orbitals is thus larger than the typical values of in-plane hopping $|t^{(2)}_{zx,zx}|\sim \qty{260}{\milli\electronvolt}$, leading to the one-dimensional electronic structures along the $c$-axis as shown in Fig.~\ref{fig:FS}.
The moderately large out-of-plane hopping between $d_{yz}$ and $d_{v}$-orbitals $|t^z_{yz,v}|~\sim\qty{112} {\milli\electronvolt}$ also brings the one-dimensionality in the $d_v$-orbital.

\section{Summary and Discussions}

\begin{table}[b]
    \centering
    \caption{Several hopping parameters with spin-up components in the $t_{2g}$ and $t_{2g}$-$p$ models in units of $\unit{\milli\electronvolt}$.}
    \begin{tabular}{r r r r }
        \hline\hline
        $t_{2g}$ model     & $t^{(1)}$ & $t^{(2)}$  & $t^z$     \\ \hline
        $zx$-$zx$          & $151+0i$  & $260-8i$   & $168-6i$  \\
        $zx$-$yz$          & $0-12i$   & $ 200+12i$ & $-399-4i$ \\
        $zx$-$v$           & $90+0i$   & $-104-16i$ & $39+0i$   \\
        $yz$-$zx$          & $0+12i$   & $-200-12i$ & $399+4i$  \\
        $yz$-$yz$          & $28+0i$   & $-232+13i$ & $328-2i$  \\
        $yz$-$v$           & $0-2i$    & $-18+26i$  & $112-7i$  \\
        $v$-$zx$           & $90+0i$   & $-104-16i$ & $39+0i$   \\
        $v$-$yz$           & $0+2i$    & $18-26i$   & $-112+7i$ \\
        $v$-$v$            & $-44+0i$  & $522-29i$  & $-9-10i$  \\
        \hline\hline
        $t_{2g}$-$p$ model & $t^{(1)}$ & $t^{(2)}$  & $t^z$     \\ \hline
        $zx$-$zx$          & $103+0i$  & $81+1i$    & $147+1i$  \\
        $zx$-$yz$          & $0-1i$    & $104+3i$   & $-146+0i$ \\
        $zx$-$v$           & $67+0i$   & $-51-4i$   & $29+2i$   \\
        $yz$-$zx$          & $0+1i$    & $-104-3i$  & $146+0i$  \\
        $yz$-$yz$          & $220+0i$  & $-105+6i$  & $103+0i$  \\
        $yz$-$v$           & $0-8i$    & $-16+13i$  & $48-2i$   \\
        $v$-$zx$           & $67+0i$   & $-51-4i$   & $29+2i$   \\
        $v$-$yz$           & $0+8i$    & $16-13i$   & $-48+2i$  \\
        $v$-$v$            & $-440+0i$ & $126-32i$  & $8-2i$    \\
        \hline\hline
    \end{tabular}
    \label{tab:hopping}
\end{table}

We presented a detailed first-principles investigation of the electronic structure of PbRe$_2$O$_6$, a promising spin-orbit-coupled metal that exhibits successive structural transitions. We showed that the the nearly despersionless bands near the Fermi level originate from the $E_g$ molecular orbitals formed by the Re-$d_{x^2-y^2}$ states, and that these flat bands are responsible for the large density of states at the Fermi level. In contrast, the quasi-one-dimensional Fermi surfaces arise from the anisotropic dispersive bands associated with the Re $d_{yz}$ and $d_{zx}$ orbitals. These characteristic features are possible sources of electronic and/or electron-lattice instabilities such as Peierls transitions, charge-density-wave transition, multipolar ordering, and structural transitions, and therefore our analysis provides insight into the origin of the successive phase transitions in PbRe$_2$O$_6$.
Regarding the structural transitions, previous first-principles calculations found the phonon instabilities in the related material PbNb$_2$O$_6$~\cite{Olsen_Inorganic_Chemistry_53.9715}.
The theoretical study on phonons in PbRe$_2$O$_6$ is a possible future work.

\begin{acknowledgments}
    The authors thank Z. Hiroi and D. Hirai for fruitful discussions.
    This research was supported by JSPS KAKENHI Grants Numbers
    JP23K20824, JP23K25827, JP24K00581, JP24K00588, JP25K21684, JP25K00947.
    Parts of numerical calculations have been performed using MASAMUNE- and MASAMUNE II-IMR at Center for Computational Materials Science, Institute for Materials Research, Tohoku University.
    Figures of crystal structures and Fermi surfaces are drawn by using \textsc{vesta}~\cite{Momma:db5098} and \textsc{FermiSurfer}~\cite{KAWAMURA2019197}, respectively.

\end{acknowledgments}

\bibliography{ref}

\end{document}